\documentclass[journal]{IEEEtran}
\usepackage{float}
\usepackage{graphicx}
\usepackage{subfigure}
\usepackage{epstopdf}
\usepackage{color}
\usepackage{amssymb}
\usepackage{amsmath}
\usepackage{algorithm}
\usepackage{algorithmic}
\usepackage{bm}
\usepackage[colorlinks,citecolor=blue,linkcolor=blue,urlcolor=blue]{hyperref}
\usepackage{stfloats}
\begin{document}
\title{\LARGE FAS for Secure and Covert Communications}
\author{Junteng Yao,
            Liangxiao Xin,
            Tuo Wu,
            Ming Jin, \emph{Member, IEEE},\\
            Kai-Kit Wong, \emph{Fellow, IEEE},
            Chau Yuen, \emph{Fellow, IEEE}
            and Hyundong Shin, \emph{Fellow}, \emph{IEEE}
\vspace{-10mm}

\thanks{J. Yao, L. Xin, and M. Jin are with the faculty of Electrical Engineering and Computer Science, Ningbo University, Ningbo 315211, China (E-mail: $\rm\{yaojunteng, xinliangxiao, jinming\}@nbu.edu.cn$). T. Wu and C. Yuen are with the School of Electrical and Electronic Engineering, Nanyang Technological University, 639798, Singapore (E-mail: $\rm\{tuo.wu, chau.yuen\}@ntu.edu.sg$). K. K. Wong is with the Department of Electronic and Electrical Engineering, University College London, Torrington Place, WC1E 7JE, United Kingdom and also with the Department of Electronic Engineering, Kyung Hee University, Yongin-si, Gyeonggi-do 17104, Korea (E-mail: $\rm kai\text{-}kit.wong@ucl.ac.uk$). H. Shin is with the Department of Electronic Engineering, Kyung Hee University, Yongin-si, Gyeonggi-do 17104, Korea (E-mail: $\rm hshin@khu.ac.kr$).}

}
\maketitle

\begin{abstract}
This letter considers a fluid antenna system (FAS)-aided secure and covert communication system, where the transmitter adjusts multiple fluid antennas' positions to achieve secure and covert transmission under the threat of an eavesdropper and the detection of a warden. This letter aims to maximize the secrecy rate while satisfying the covertness constraint. Unfortunately, the optimization problem is non-convex due to the coupled variables. To tackle this, we propose an alternating optimization (AO) algorithm to alternatively optimize the optimization variables in an iterative manner. In particular, we use a penalty-based method and the majorization-minimization (MM) algorithm to optimize the transmit beamforming and fluid antennas' positions, respectively. Simulation results show that FAS can significantly improve the performance of secrecy and covertness compared to the fixed-position antenna (FPA)-based schemes.
\end{abstract}


\vspace{-2mm}
\section{Introduction}
\IEEEPARstart{W}{ith the} development of internet-of-things (IoT) networks, wireless communication has brought great convenience to our lives \cite{IYaqoob17}. However, the broadcasting nature of wireless signals raises significant security concerns, making it vulnerable to eavesdropping. To tackle this issue, physical layer security (PLS) has emerged as an effective technique to prevent legitimate information from being intercepted by eavesdroppers \cite{BZhao23,QLi2023}. In addition to eavesdropping concerns, another critical challenge is the need for covert communication, where the transmission behavior itself must be concealed from detection \cite{JHu24,CWang23}. Covert communication aims to ensure that the transmission remains undetectable to wardens. As a result, researchers have extensively explored and investigated secure and covert communication techniques in different wireless systems, e.g., \cite{QLi23,HWu24,MForouzesh20}, to name a few.

Nevertheless, the aforementioned works primarily focus on transceivers equipped with fixed-position antennas (FPAs) that limit their ability to fully exploit spatial diversity gains. To overcome this limitation, fluid antenna systems (FASs) have emerged as a promising solution \cite{KKWong21,Wong-2022fcn,New-submit2024,Yao20241}. Specifically, FAS can reconfigure antenna positions dynamically, selecting the optimal locations in real-time to provide higher degrees of freedom (DoFs) and maximize spatial resource utilization. Inspired by these capabilities, FAS has been adopted for many applications, including channel estimation \cite{Hao-2024}, beamforming \cite{New-twc2023,Wang-aifas2024}, and resource allocation \cite{JYao24,CWang24,LZhou24,JTang24}.

Encouraged by these advantages, integrating FAS into secure and covert communication systems can offer great security benefits. A FAS-enabled transmitter can engineer channel conditions across all links by adjusting the positions of the antennas, for not only reducing the probability of detection by potential wardens but improving the secrecy rate for legitimate users. Nonetheless, the impact of FAS on secure, covert communication systems is not known. Effectively leveraging FAS requires addressing challenging optimization problems, such as obtaining the optimal antenna positions and beamforming vectors to maximize secrecy rate while under covert.

Motivated by the above, this letter considers the adoption of FAS at the transmitter in secure and covert communication systems, in which the transmitter sends signals to a legitimate user under the threat of a potential eavesdropper and detection of a warden. Our goal is to maximize the secrecy rate of the legitimate user while satisfying the detection error probability (DEP) constraints of the warden and limiting the transmitter's power consumption. The optimization problem is non-convex due to the coupled optimizing variables. To solve it efficiently, we propose an alternating optimization (AO) approach. Specifically, for the transmit beamforming problem, we utilize a penalty-based scheme to manage the rank-one constraint while ensuring convergence. For the antenna position optimization sub-problem, we employ the majorization-minimization (MM) algorithm. Simulation results demonstrate that the proposed scheme significantly outperforms existing benchmarks in terms of secure and covert communication performance.


\section{ System Model and Problem Formulation}
We consider a FAS-aided secure and covert communication system, where a transmitter (Alice) equipped with $N (N\geq 2)$ fluid antennas transmits signals to the legitimate user (Bob) equipped with a single FPA, an eavesdropper (Eve) equipped with a single FPA overhears the signals, and a warden (Willie) equipped with a single FPA tries to detect if a transmission from Alice exists. Alice adjusts the positions of fluid antennas, which are connected to $N$ radio frequency (RF) chains via integrated waveguides \cite{Zhang-pFAS2024} or flexible cables, in a finite range $\mathcal{S}_t$ to achieve secure and covert communications. We adopt the planar far-field response model to express the channels. In this case, adjusting the antennas positions will change the phase of path response coefficients, while not influencing the angles of arrival/departure (AoAs/AoDs), and the amplitude of path response coefficients for each channel path component \cite{JTang24}. Denote the $n$-th fluid antenna's position as $\mathbf{t}_{n}=[x_n, y_n]^T$, and the positions set in Alice is $\mathbf{\bar{t}} = [\mathbf{t}_1, \dots, \mathbf{t}_N]$.

We assume that the number of transmit paths in Alice-Bob link, Alice-Eve link, and Alice-Willie link are $L_b^t$, $L_e^t$ and $L_w^t$, respectively. For the two links, the propagation distance difference between the $n$-th fluid antenna and the reference origin in the $l$-th transmit path as $\rho_{k,l}(\mathbf{t}_n) = x_{n}^{t}\sin\theta_{k,l}^t \cos\varphi_{k,l}^t + y_{n}^{t} \cos\theta_{k,l}^t$, where $k\in\{b,e,w\}, l\in\mathcal{L}=\{1,\dots, L\}$, $\theta_{k,l}^t \in [0, \pi]$ and $\varphi_{k,l}^t \in [0, \pi]$ represent the elevation and azimuth angles of the $l$-th path, respectively. Thus, the transmit field response vectors of the $n$-th fluid antenna in the Alice-Bob link, Alice-Eve link, and Alice-Willie link are given by
\begin{align}\label{eq1}
\mathbf{f}_k(\mathbf{t}_n)=& \left [e^{j \frac{2\pi}{\lambda}\rho_{k,1}(\mathbf{t}_n)}, \dots, e^{j \frac{2\pi}{\lambda}\rho_{k,L_k^t}(\mathbf{t}_n)}\right]^T
\end{align}
for $k\in\{b,e,w\}$, where $\lambda$ is the carrier wavelength. Thus, the transmit field response matrices of the Alice-Bob link, Alice-Eve link, and Alice-Willie link are expressed as
\begin{align}\label{eq2}
\mathbf{F}_k\mathbf{(\bar{t})} =& \left [\mathbf{f}_k(\mathbf{t}_1),\mathbf{f}_k(\mathbf{t}_2), \dots, \mathbf{f}_k(\mathbf{t}_N)\right], k\in\{b,e,w\},
\end{align}

Furthermore, we define the path response matrices $\bm{\Sigma}_b\in\mathbb{C}^{{L_b^r}\times {L_b^t}}$, $\bm{\Sigma}_e\in\mathbb{C}^{{L_e^r}\times {L_e^t}}$, and $\bm{\Sigma}_w\in\mathbb{C}^{{L_w^r}\times {L_w^t}}$ as the path responses of the Alice-Bob link, Alice-Eve link, and Alice-Willie link, respectively. Considering that a single FPA exists in Bob, Eve, and Willie, the channel of the Alice-Bob link and Alice-Willie link can be written as
\begin{align}\label{eq4}
\mathbf{h}_k^H=&\mathbf{1}^H\bm{\Sigma}_k\mathbf{F}_k\mathbf{(\bar{t})}\in\mathbb{C}^{1\times N}, k\in\{b,e,w\}.
\end{align}

Alice transmits the signal $x$ with $\mathbb{E}\left(|x|^2 \right)=1$ to Bob and the signal-to-noise ratios (SNRs) of Bob and Eve are found as
\begin{align}\label{eq5}
\gamma_k=\mathrm{Tr}\left(\mathbf{H}_k\mathbf{V}\right)/\sigma^2, k\in\{b,e\},
\end{align}
where $\mathbf{H}_k=\mathbf{h}_k\mathbf{h}_k^H$, $\mathbf{V}=\mathbf{v}\mathbf{v}^H$, $\mathbf{v} \in \mathbb{C}^{N \times 1}$ is the transmit beamforming vector, $\sigma^2$ represents the noise power. As such, the secrecy rate under weak secrecy conditions is given by
\begin{align}\label{eq6}
R_s=\log_2\left(1+\gamma_b\right)-\log_2\left(1+\gamma_e\right).
\end{align}

From the perspective of Willie, the received signal at Willie can be expressed as
\begin{equation}\label{eq8}
y_w=\left\{ \begin{array}{lc}
n_w, &  \mathcal{H}_0,\\
\mathbf{g}^H\mathbf{v}x+n_w, & \mathcal{H}_1,
\end{array}\right.
\end{equation}
where $\mathcal{H}_0$ denotes the null hypothesis that no transmission has occurred, $\mathcal{H}_1$ represents the alternative hypothesis that Alice has transmitted, and $n_w\sim \mathcal{CN}(0, \sigma^2)$ is the complex additive white Gaussian noise (AWGN) at Willie.

Denoting $p_0(y_w)=f(y_w|\mathcal{H}_0)$ and $p_1(y_w)=f(y_w|\mathcal{H}_1)$ as the likelihood functions of $y_w$ under $\mathcal{H}_0$ and $\mathcal{H}_1$, we have
\begin{align}\label{eq9}
p_i(y_w)=\frac{1}{\pi\delta_i}\exp(-|y_w|^2/\delta_i), i\in\{0, 1\},
\end{align}
where $\delta_0=\sigma^2$, $\delta_1=\mathrm{Tr}\left(\mathbf{H}_w\mathbf{V}\right)+\sigma^2$, and $\mathbf{H}_w=\mathbf{h}_w\mathbf{h}_w^H$. The priori probabilities of the hypotheses are assumed equal.

For Willie, the detection performance, i.e., DEP, can be found as $\varepsilon=\mathbb{P}(D_1|\mathcal{H}_0)+\mathbb{P}(D_0|\mathcal{H}_1)$, where $D_0$ indicates that Alice dose not transmit signals, $D_1$ denotes the other case, and $\mathbb{P}(D_1|\mathcal{H}_0)$  and $\mathbb{P}(D_0|\mathcal{H}_1)$ are the false alarm probability and the miss detection probability, respectively.

Due to the fact that the computation of $\varepsilon$ is intractable, we obtain a tractable lower bound of $\varepsilon$ by using Pinsker's inequality \cite{JHu24}, which is given by $\varepsilon\geq 1-\sqrt{\frac{1}{2}\mathbb{D}(p_0\|p_1)}$, where $\mathbb{D}(p_0\|p_1)$ denotes the Kullback-Leibler  divergence from $p_0$ to $p_1$, and its exact expression is $\mathbb{D}(p_0\|p_1)=\ln\frac{\sigma_1}{\sigma_0}+\frac{\sigma_0}{\sigma_1}-1$.

Given a predetermined tolerated detection coefficient $\epsilon\in[0,1]$, to ensure successful covert transmission, $\mathbb{D}(p_0\|p_1)$ should satisfy the following condition \cite{CWang23}:
\begin{align}\label{eq13}
\mathbb{D}(p_0\|p_1)\leq 2\epsilon^2.
\end{align}
Defining $a\triangleq \delta_1/\delta_0$, \eqref{eq13} can be reformulated as $f(a)\triangleq\ln a+\frac{1}{a}\leq 1+2\epsilon^2$. By introducing $a_1$ and $a_2$ as the two roots of $f(a)=1+2\epsilon^2$, we can obtain the range of $a$, i.e., $a_1\leq a \leq a_2$, where $a_1=\exp(\mathcal{W}_{-1}(-\exp(-(1+2\epsilon^2)))+1+2\epsilon^2)$, $a_2=\exp(\mathcal{W}_{0}(-\exp(-(1+2\epsilon^2)))+1+2\epsilon^2)$, and $\mathcal{W}(z)$ denotes the Lambert $\mathcal{W}$ function. Since $a=1+\mathrm{Tr}\left(\mathbf{H}_w\mathbf{V}\right)/\sigma^2>1$, we have
\begin{align}\label{eq14}
\mathrm{Tr}\left(\mathbf{H}_w\mathbf{V}\right)\leq \sigma^2(a_2-1).
\end{align}

We aim to maximize the secrecy rate of Bob while satisfying the constraints on Alice' transmit power, the fluid antennas' positions, and the covertness requirement. After removing the log due to monotonicity, the optimization problem becomes
\begin{subequations}\label{eq15}
\begin{align}
\max\limits_{\mathbf{\bar{t}},\mathbf{V}} \quad \ & \frac{1+\gamma_b}{1+\gamma_e} \label{eq15a}\\
\mathrm{s.t.} \quad \  &\mathbf{\overline{t}} \in \mathcal{S}_t, \label{eq15b}\\
&||\mathbf{t}_n-\mathbf{t}_v||_2\geq D,~n,v\in\mathcal{N},~n\neq v, \label{eq15c}\\
&\mathrm{Tr}(\mathbf{V}) \leq P_{\max}, \label{eq15d}\\
&\mathrm{rank}(\mathbf{V})=1, \label{eq15e}\\
&\eqref{eq14},
\end{align}
\end{subequations}
where \eqref{eq15c} corresponds to the minimum distance requirement between any two antennas within the transmit region, \eqref{eq15d} denotes the maximum transmit power constraint of Alice, and \eqref{eq15e} is the rank-one constraint of $\mathbf{V}$. Due to the objective function and constraints being non-convex, \eqref{eq15} is non-convex. To proceed, we introduce two auxiliary variables, i.e., $\beta_1$ and $\beta_2$, and Problem \eqref{eq15} can be relaxed to
\begin{subequations}\label{eq151}
\begin{align}
\max\limits_{\mathbf{\bar{t}},\mathbf{V},\beta_1>,\beta_2>0} \quad \ & \beta_1 \label{eq151a}\\
\mathrm{s.t.} \quad \  &\sigma^2+\mathrm{Tr}\left(\mathbf{H}_b\mathbf{V}\right)\geq\beta_1\beta_2, \label{eq151b}\\
&\beta_2\geq\sigma^2+\mathrm{Tr}\left(\mathbf{H}_e\mathbf{V}\right), \label{eq151c}\\
&\eqref{eq14}, \eqref{eq15c}\text{--}\eqref{eq15e}.
\end{align}
\end{subequations}
As Problem \eqref{eq151} is still non-convex, in the subsequent section, we employ an AO algorithm to tackle this problem.

\vspace{-2mm}
\section{AO Algorithm}
In this section, we use the AO algorithm to decompose Problem \eqref{eq151} into $N+1$ sub-problems, and alternately optimize these sub-problems to obtain a locally optimal solution.

\subsection{Alice's Transmit Beamforming Optimization}
Given $\mathbf{\bar{t}}$, Problem \eqref{eq151} can be reexpressed as
\begin{align}\label{eq16}
\max\limits_{\mathbf{V},\beta_1>0,\beta_2>0} \  \beta_1 \quad \mathrm{s.t.} \ \eqref{eq14},\eqref{eq15d}, \eqref{eq15e}, \eqref{eq151b}, \eqref{eq151c},
\end{align}
which is still a non-convex optimization problem due to the non-convex constraints \eqref{eq15e} and \eqref{eq151b}.

For the non-convex rank-one constraint \eqref{eq15e}, we propose a penalty-based algorithm to overcome this issue. Specifically, for a positive semidefinite matrix $\mathbf{V}$, its trace and the largest eigenvalue satisfy $\mathrm{Tr}(\mathbf{V})-\lambda_\text{max}\left( \mathbf{V}\right)\geq0$, and the equality holds when $\mathrm{rank}(\mathbf{V})=1$. Furthermore, because $\lambda_\text{max}\left( \mathbf{V}\right)$ is not differentiable, we use $\mathbf{u}_{\text{max},m}^H\mathbf{V}\mathbf{u}_{\text{max},m}$ to approximate $\lambda_\text{max}\left( \mathbf{V}\right)$, where $\mathbf{u}_{\text{max},m}$ is the eigenvector corresponding to the largest eigenvalue $\lambda_\text{max}\left( \mathbf{V}^{(m)}\right)$ in the $m$-th iteration.

For $\beta_1\beta_2$ in the constraint \eqref{eq151b}, we have
\begin{align}
\beta_1\beta_2=\frac{1}{4}\left(\beta_1+\beta_2\right)^2-\frac{1}{4}\left(\beta_1-\beta_2\right)^2.
\end{align}
Note that $\frac{1}{4}\left(\beta_1+\beta_2\right)^2$ and $\frac{1}{4}\left(\beta_1-\beta_2\right)^2$ are convex on both $\beta_1$ and $\beta_2$. Thus, we have the upper bound of $\beta_1\beta_2$ as
\begin{multline}
f(\beta_1, \beta_2;\beta_1^{(m)}, \beta_2^{(m)})= \frac{1}{4}\left(\beta_1+\beta_2\right)^2-\frac{1}{4}\left(\beta_1^{(m)}-\beta_2^{(m)}\right)^2\\
-\frac{1}{2}\left(\beta_1^{(m)}-\beta_2^{(m)}\right)\left(\beta_1-\beta_1^{(m)}-\beta_2+\beta_2^{(m)}\right).
\end{multline}
Accordingly, the constraint \eqref{eq151b} can be relaxed to
\begin{align}\label{eq00}
\sigma^2+\mathrm{Tr}\left(\mathbf{H}_b\mathbf{V}\right)\geq f(\beta_1, \beta_2;\beta_1^{(m)}, \beta_2^{(m)}).
\end{align}

Therefore, Problem \eqref{eq16} can be reformulated as
\begin{subequations}\label{eq17}
\begin{align}
\max\limits_{\mathbf{V},\beta_1>0,\beta_2>0} \ \ & \beta_1 -\eta \left(\mathrm{Tr}(\mathbf{V})-\mathbf{u}_{\text{max},m}^H\mathbf{V}\mathbf{u}_{\text{max},m}\right)\label{eq15a}\\
\mathrm{s.t.}  \  \ &\eqref{eq14}, \eqref{eq15d},\eqref{eq151c},\eqref{eq00},
\end{align}
\end{subequations}
where $\eta$ is a penalty factor. Now, \eqref{eq17} is convex and it can be solved using the convex programming toolbox CVX \cite{MGrant}.

\subsection{The $n$-th Fluid Antenna's Position Optimization}
Given $\mathbf{V}$ and $\{\mathbf{t}_v\}_{v\neq n}$, Problem \eqref{eq15} is formulated as
\begin{align}\label{eq18}
\max\limits_{\mathbf{t}_n,\beta_1>0,\beta_2>0} \ \beta_1  \quad \mathrm{s.t.} \ \eqref{eq14}, \eqref{eq15b}, \eqref{eq15c}, \eqref{eq151b}, \eqref{eq151c}.
\end{align}
Problem \eqref{eq18} is non-convex due to the non-convex constraints. To solve $\{\mathbf{t}_m\}_{m\neq n}$, we first rewrite $\mathrm{Tr}\left(\mathbf{H}_k\mathbf{V}\right)$ as
\begin{multline}\label{eq21}
\mathrm{Tr}\left(\mathbf{1}^H\bm{\Sigma}_k\mathbf{F}_k\mathbf{(\bar{t})} \mathbf{v} \mathbf{v}^H
\mathbf{F}_k^H\mathbf{(\bar{t})} \bm{\Sigma}_k^H \mathbf{1} \right)\\
=\alpha_k+\beta_b(\mathbf{t}_n)+2\mathbb{R}\{\mathbf{f}_k^H(\mathbf{t}_n) \bm{\Omega}_k\},k\in\{b,e,w\},
\end{multline}
where $\alpha_k=\mathrm{Tr}\left(\sum_{j\neq n}^{N} \mathbf{f}_k(\mathbf{t}_j)\mathbf{v}(j) \sum_{l\neq n}^{N} \mathbf{v}^H(l)\mathbf{f}_k^H(\mathbf{t}_l)\bm{\Phi}_k \right)$, $\beta_k(\mathbf{t}_n)=\mathrm{Tr}\left( \mathbf{v}(n)\mathbf{v}^H(n)\mathbf{f}_k(\mathbf{t}_n)\mathbf{f}_k^H(\mathbf{t}_n)\bm{\Phi}_k  \right)$, $\bm{\Phi}_k=\bm{\Sigma}_k^H\bm{\Sigma}_k$, and $\bm{\Omega}_k=\bm{\Phi}_k\left(\sum_{j\neq n}^{N} \mathbf{f}_k(\mathbf{t}_j)\mathbf{v}(j)\right)\mathbf{v}^H(n)$.

Then we first obtain a lower bound of $\beta_k(\mathbf{t}_n)$ through the first-order Taylor expansion at point $\mathbf{f}_k(\mathbf{t}_n^{(m)})$, which is
\begin{align}\label{eq26}
2\mathbb{R}\{ \mathbf{f}_k^H(\mathbf{t}_n^{(m)})  \bm{\Psi}_k \mathbf{f}_k(\mathbf{t}_n) \}- \mathbf{f}_k^H(\mathbf{t}_n^{(m)}) \bm{\Psi}_k \mathbf{f}_k(\mathbf{t}_n^{(m)}),
\end{align}
where $\mathbf{t}_n^{(m)}$ is $\mathbf{t}_n$ at the $m$-th iteration, $\bm{\Psi}_k=\bm{\Phi}_k\mathbf{v}(n)\mathbf{v}^H(n)$. Now, we combine the third term in \eqref{eq21} with the first term in \eqref{eq26}, and thus have $\bar{\beta}_k(\mathbf{t}_n)=2\mathbb{R}\{\mathbf{f}_k^H(\mathbf{t}_n) \bm{\Upsilon}^n_k\}$, where $\bm{\Upsilon}^n_k=\bm{\Psi}_k^H \mathbf{f}_k(\mathbf{t}_n^{(m)})+\bm{\Omega}_k$. We further use the second-order Taylor expansion to construct a global lower bound of $\bar{\beta}_k(\mathbf{t}_n)$ as
\begin{multline}\label{eq28}
g_k^l(\mathbf{t}_n) ={\bar{\beta}_k}(\mathbf{t}^{(m)}_n)+\nabla {\bar{\beta}_k}(\mathbf{t}_n^{(m)})^T\left(\mathbf{t}_n-\mathbf{t}_n^{(m)}\right)\\
-\frac{\kappa_k^n}{2}\left(\mathbf{t}_n-\mathbf{t}_n^{(m)}\right)^T\left(\mathbf{t}_n-\mathbf{t}_n^{(m)}\right),
\end{multline}
where $\kappa_k^n=\frac{16\pi^2}{\lambda^2} \sum_{l=1}^{L_k^t} \lvert \bm{\Upsilon}_k^{l} \rvert$, and $\nabla {\bar{\beta}_k}(\mathbf{t}_n)$ is the gradient vector of $\bar{\beta}_k(\mathbf{t}_n)$, with detailed derivations given in Appendix \ref{appendixA}. Therefore, a concave lower bound of $\mathrm{Tr}\left(\mathbf{H}_k\mathbf{V}\right)$ is
\begin{align}\label{eq29}
f^l_k\left(\mathbf{t}_n\right)=g_k^l(\mathbf{t}_n) + \alpha_k - \mathbf{f}_k^H(\mathbf{t}_n^{(m)}) \bm{\Psi}_k\mathbf{f}_k(\mathbf{t}_n^{(m)}).
\end{align}

Then we employ the MM algorithm to obtain the upper bound of $\mathrm{Tr}\left(\mathbf{H}_k\mathbf{V}\right), k\in\{b,e,w\}$. According to \cite{JTang24}, for a given $\mathbf{f}_k^H(\mathbf{t}_n^{(m)}), k\in\{e,w\}$, the following inequality holds for any feasible $\mathbf{f}_k^H(\mathbf{t}_n)$, i.e.,
\begin{multline}\label{eq37}
\beta_k(\mathbf{t}_n) \leq \mathbf{f}_k^H(\mathbf{t}_n) \bm{\Theta}_k \mathbf{f}_k(\mathbf{t}_n) + \mathbf{f}_k^H(\mathbf{t}_n^{(m)}) \left( \bm{\Theta}_k - \bm{\Psi}_k\right) \mathbf{f}_k(\mathbf{t}_n^{(m)})\\
-2\mathbb{R}\left\{\mathbf{f}_k^H(\mathbf{t}_n) \left( \bm{\Theta}_k - \bm{\Psi}_k\right) \mathbf{f}_k(\mathbf{t}_n^{(m)})\right\},
\end{multline}
where $\bm{\Theta}_k=\lambda^k_\text{max}\mathbf{I}_{L_k^t}$, and $\lambda^k_\text{max}$ is the maximum eigenvalue of $\bm{\Psi}_k$. From \eqref{eq37}, we know that $\mathbf{f}_k^H(\mathbf{t}_n) \bm{\Theta}_k \mathbf{f}_k(\mathbf{t}_n)=\lambda^k_\text{max} L_k^t$, and the second term is constant. Then we define
\begin{align}\label{eq40}
c_k(\mathbf{t}_n)\triangleq2\lambda^k_\text{max} L_k^t-\mathbf{f}_k^H(\mathbf{t}_n^{(m)}) \bm{\Psi}_k \mathbf{f}_k(\mathbf{t}_n^{(m)}).
\end{align}
The third term in \eqref{eq21} can be combined with the third term in \eqref{eq37}, and we get $\tilde{\beta}_k(\mathbf{t}_n)=2\mathbb{R}\{\mathbf{f}_k^H(\mathbf{t}_n) \bm{\Pi}_k^n\}$, where $\bm{\Pi}_k^n=\bm{\Omega}_k- \left( \bm{\Theta}_k - \bm{\Psi}_k \right) \mathbf{f}_k(\mathbf{t}_n^{(m)})$. Then we also utilize the second-order Taylor expansion to construct the upper bound of $\tilde{\beta}_k(\mathbf{t}_n)$, which is given by
\begin{multline}\label{eq42}
g_k^u(\mathbf{t}_n) ={\tilde{\beta}_k}(\mathbf{t}_n^{(m)})+\nabla {\tilde{\beta}_k}(\mathbf{t}_n^{(m)})^T\left(\mathbf{t}_n-\mathbf{t}_n^{(m)}\right)\\
+\frac{\tilde{\kappa}_k^n}{2}\left(\mathbf{t}_n-\mathbf{t}_n^{(m)}\right)^T\left(\mathbf{t}_n-\mathbf{t}_n^{(m)}\right),
\end{multline}
where $\tilde{\kappa}_k^n=\frac{16\pi^2}{\lambda^2} \sum_{l=1}^{L_k^t} \lvert \bm{\Pi}_k^{l} \rvert$, and $\nabla {\tilde{\beta}_k}(\mathbf{t}_n^{(m)})$ is similarly obtained with the derivations in Appendix \ref{appendixA}. Therefore, we can obtain the upper bound of $\mathrm{Tr}\left(\mathbf{H}_k\mathbf{V}\right)$ as
\begin{align}\label{eq48}
f^u_k\left(\mathbf{t}_n\right)=g_k^u(\mathbf{t}_n^{(m)})  + \alpha_k +c_k(\mathbf{t}_n).
\end{align}

Based on the above-mentioned derivations, the constraint \eqref{eq14}, \eqref{eq151b}, and \eqref{eq151c} can be relaxed as
\begin{align}
\label{eq49a}f_w^u(\mathbf{t}_n) \leq& \sigma^2(a_2-1),\\
\label{eq49b}\sigma^2+f^l_b\left(\mathbf{t}_n\right)\geq& f(\beta_1, \beta_2;\beta_1^{(m)}, \beta_2^{(m)}),\\
\label{eq49c}\beta_2\geq&\sigma^2+f_e^u(\mathbf{t}_n).
\end{align}

For \eqref{eq15c}, we can relax $\|\mathbf{t}_n-\mathbf{t}_v\|_2$ to be a concave function of $\mathbf{t}_n$ as its lower bound using the first-order Taylor expansion at point $\mathbf{t}_n^{(m)}$. Then the constraint \eqref{eq15c} can be written as
\begin{align}\label{eq51}
\frac{1}{\|\mathbf{t}_n^{(m)}-\mathbf{t}_v\|_2}(\mathbf{t}_n^{(m)}-\mathbf{t}_v)^T(\mathbf{t}_n-\mathbf{t}_v) \geq D.
\end{align}

Therefore, Problem \eqref{eq18} can be reformulated as
\begin{align}\label{eq52}
\max\limits_{\mathbf{t}_n,\beta_1>0,\beta_2>0} \  \beta_1 \quad \mathrm{s.t.} \ \eqref{eq15b},\eqref{eq49a}\text{--}\eqref{eq51},
\end{align}
which is convex, and can also be solved using CVX \cite{MGrant}.

\section{Simulation Results}
In the simulations, we consider a two-dimensional coordinate system, where Alice, Bob, Eve, and Willie are located at (0,0) m, (100,0) m, (150,5) m, and (150,-5) m, respectively. We assume that the carrier frequency is 2.4 GHz with a wavelength $\lambda$ = 0.125 m, and a minimum inter-antenna distance is $D=\frac{\lambda}{2}$. We assume that the spatial region of fluid antennas at Alice is $\mathcal{S}_t=\left[-{\frac{A}{2}},{\frac{A}{2}}\right] \times \left[-{\frac{A}{2}},{\frac{A}{2}}  \right]$, where $A=4\lambda$. Besides, the number of transmit paths equals to the number of receive paths for all links, i.e., $L_k^t=L_k^r=L=4, k\in\{b,e,w\}$. The path response matrices of all links are modeled as $\bm{\Sigma}_k[l,l] \sim \mathcal{CN}(0,g_0  d_k^{-\alpha}/L), l\in\mathcal{L}$, where $d_k$ is the distance, $g_0=-40$ dB represents the average channel gain at the reference distance 1 m and $\alpha=2.8$ is the path loss coefficient. The maximum transmit power of Alice is $P_\text{max}=20$ dBm, the noise power $\sigma^2=-80$ dBm, and the tolerated detection coefficient is $\epsilon=0.2$. In the transmit beamforming optimization, the initial penalty factor $\eta=1$, and $\eta\rightarrow1.5\eta$ in each interior iteration. Finally, the convergence accuracy is set to $10^{-4}$.

\begin{figure}[t]
\centering
\includegraphics[width=2.8in]{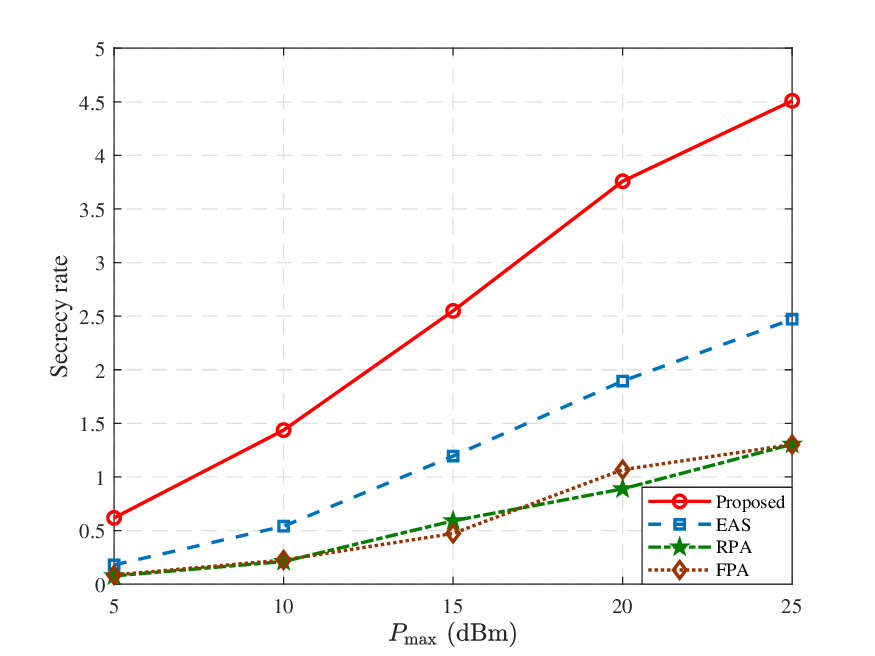}
\caption {The maximum transmit power of Alice $P_{\max}$ versus the secrecy rate.}\label{SRvsP}
\vspace{-4mm}
\end{figure}

\begin{figure}[t]
\centering
\includegraphics[width=2.8in]{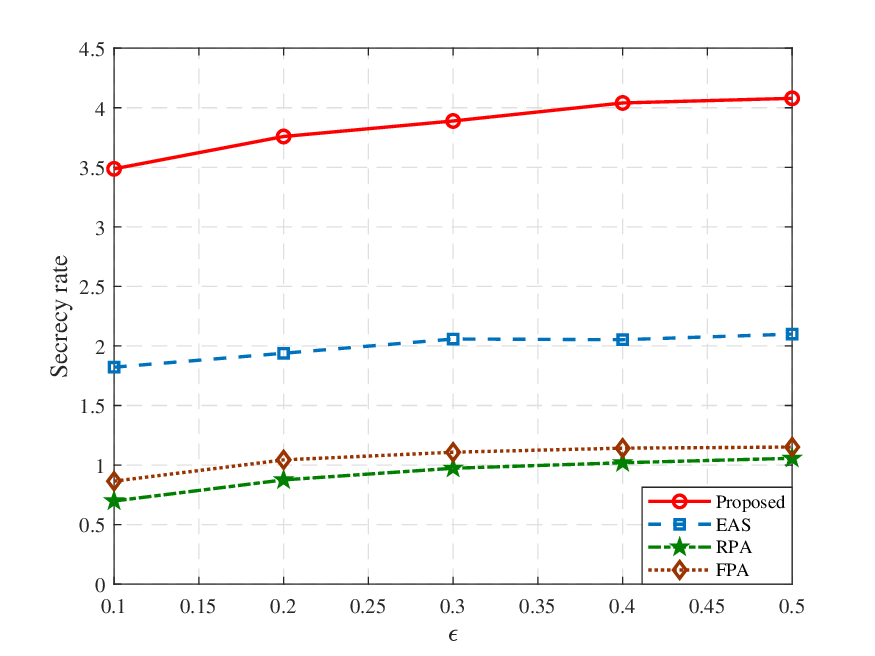}
\caption {The tolerated detection coefficient $\epsilon$ versus the secrecy rate.}\label{SRvsepsilon}
\vspace{-4mm}
\end{figure}

In Fig.~\ref{SRvsP}, the results are provided to investigate the impact of Alice's maximum transmit power on the secrecy rate. In the legend, ``Proposed'' represents our proposed FAS scheme, while ``FPA'' denotes the scheme that $N$ FPAs are spaced at a spacing of $\lambda/2$. Also, ``RPA'' is the scheme that $N$ antennas are randomly distributed within the transmit region $\mathcal{S}$, satisfying the constraint in \eqref{eq15c}, and ``EAS'' has all $N$ antennas selected from $2N$ fixed positions using an exhaustive search approach. From Fig.~\ref{SRvsP}, we observe that with the increasing of $P_{\max}$, the secrecy rate enhances in all the schemes. Moreover, our proposed scheme consistently outperforms other schemes.

Fig.~\ref{SRvsepsilon} shows the results to analyze the relationship between the tolerated detection coefficient $\epsilon$ and the secrecy rate. Our proposed scheme consistently demonstrates superior performance compared to the benchmark schemes. As $\epsilon$ increases, the secrecy rate also rises for all the schemes. This is because a higher $\epsilon$ relaxes the constraint in \eqref{eq14}, allowing more spatial resource to be allocated to enhance the secrecy rate. \textit{This trend underscores the trade-off between the desired level of covert communication and the achievable secrecy rate, with our scheme effectively balancing both objectives. }

\vspace{-4mm}
\section{Conclusion}
In this letter, we studied a FAS-aided secure and covert communication system, where Alice equipped with fluid antennas sends signals to Bob while Willie attempts to detect the transmission between them and Eve overhears it. We maximized the secrecy rate of Bob under the covertness requirement by designing the transmit beamforming and antennas' positions of Alice. Simulation results verified the extraordinary impact of FAS on enhancing secure and covert communications.

\vspace{-3mm}
\appendices
\section{Derivations of $\nabla {\bar{\beta}_k}(\mathbf{t}_n)$}\label{appendixA}
To derive $\nabla {\bar{\beta}_k}(\mathbf{t}_n)$, we begin by having
\begin{align}\label{a1}
{\bar{\beta}_k}(\mathbf{t}_n)=2\left( \sum_{l=1 }^{L_k^t}   |\bm{\Upsilon}^n_k | \cos\left(\chi_k^{l}(\mathbf{t}_n)\right)\right), k\in\{b,e,w\},
\end{align}
where $\chi_k^{l}(\mathbf{t}_n)=\frac{2\pi}{\lambda}\rho_{k,l}(\mathbf{t}_n)-\angle\bm{\Upsilon}^n_k$.

As a result, the gradient vector of ${\bar{\beta}_k}(\mathbf{t}_n)$ can be respresented as $ \nabla {\bar{\beta}_k}(\mathbf{t}_n)=\left[\frac{\partial {\bar{\beta}_k}(\mathbf{t}_n)}{\partial x^{t}_n},\frac{\partial {\bar{\beta}_k}(\mathbf{t}_n)}{\partial y^{t}_n}  \right]$, given by
\begin{align}
\frac{\partial{\bar{\beta}_k}(\mathbf{t}_n)}{\partial x^{t}_n} =&  \frac{-4\pi}{\lambda}\sum_{l=1}^{L_k^t} \lvert \bm{\Upsilon}^n_k \rvert\sin\theta_{k,l}^t\cos\varphi_{k,l}^t\sin(\chi_k^{l}(\mathbf{t}_n)), \label{a3} \\
\frac{\partial{\bar{\beta}_k}(\mathbf{t}_n)}{\partial y^{t}_m} =&  \frac{-4\pi}{\lambda}\sum_{l=1}^{L_k^t} \lvert \bm{\Upsilon}^n_k \rvert \cos\phi_{k,l}^t\sin(\chi_k^{l}(\mathbf{t}_n)).  \label{a4}
\end{align}

\end{document}